\newif\ifarxiv
\arxivtrue 

\ifarxiv
\documentclass[aps,prl,reprint,nofootinbib,groupedaddress]{revtex4-1}
\usepackage{graphicx}
\else
\documentclass{nature-mod}
\usepackage{graphicx}
\makeatletter
\let\saved@includegraphics\includegraphics
\AtBeginDocument{\let\includegraphics\saved@includegraphics}
\renewenvironment*{figure}{\@float{figure}}{\end@float}
\makeatother
\fi

\graphicspath{ {./Figures/} }

\ifarxiv
\newenvironment{linenumbers}{}{}
\else
\usepackage[displaymath]{lineno}
\fi

\usepackage[utf8]{inputenc}
\usepackage{wrapfig}
\usepackage{physics}
\usepackage{setspace}
\usepackage[table]{xcolor} 
\usepackage{amsmath}
\usepackage{amsfonts}
\newcommand{\f}{\mkern-2mu f\mkern-3mu}
\newcommand\blfootnote[1]{%
  \begingroup
  \renewcommand\thefootnote{}\footnote{#1}%
  \addtocounter{footnote}{-1}%
  \endgroup
}

\ifarxiv
\let\subcite = \cite
\else
\usepackage[style=nature,url=false,isbn=false,doi=false, Month=false,date=year]{biblatex}
\DeclareFieldFormat{book}{\thefield{entrykey}}
\let\subcite = \cite
\let\cite=\supercite
\fi

\begin{document}

\ifarxiv\else
\let\oldequation\equation
\let\oldendequation\endequation

\renewenvironment{equation}
  {\linenomath\oldequation}
  {\oldendequation\endlinenomath}
  
\let\oldalign\align
\let\oldendalign\endalign
\renewenvironment{align}
  {\linenomath\oldalign}
  {\oldendalign\endlinenomath}
\fi  
  
\title{Error-corrected gates on an encoded qubit}
 
\ifarxiv
\author{Philip Reinhold$^{1,2,*,\dag}$, Serge Rosenblum$^{1,2,*\dag,\ddagger}$, Wen-Long Ma$^{1,2}$, Luigi Frunzio$^{1,2}$, Liang Jiang$^{1,2,\S}$ \& Robert J. Schoelkopf$^{1,2}$}
 \affiliation{$^1$Department of Applied Physics and Physics, Yale University, New Haven, CT, USA}
 \affiliation{$^2$Yale Quantum Institute, Yale University, New Haven, CT, USA}

\blfootnote{$\dag$ These authors contributed equally.}
\blfootnote{*\,email:\,philip.reinhold@yale.edu; serge.rosenblum@weizmann.ac.il}
\blfootnote{$\ddagger$ Current address: Department of Condensed Matter Physics, Weizmann Institute of Science, Rehovot, Israel}
\blfootnote{$\S$ Current address: Pritzker School of Molecular Engineering, University of Chicago, Chicago, Illinois 60637, USA} 
\else
\author{Philip Reinhold$^{1,2}*\dag$, Serge Rosenblum$^{1,2,3}*\dag$, Wen-Long Ma$^{1,2}$, Luigi Frunzio$^{1,2}$, Liang Jiang$^{1,2,4}$ \& Robert J. Schoelkopf$^{1,2}*$}
\maketitle
\begin{affiliations}
\blfootnote{
 \item Department of Applied Physics and Physics, Yale University, New Haven, CT, USA
 \item Yale Quantum Institute, Yale University, New Haven, CT, USA
 \item Current address: Department of Condensed Matter Physics, Weizmann Institute of Science, Rehovot, Israel
 \item Current address: Pritzker School of Molecular Engineering, University of Chicago, Chicago, Illinois 60637, USA 
 \item[] $\dag$ These authors contributed equally
 \item[] * e-mail: philip.reinhold@yale.edu; serge.rosenblum@weizmann.ac.il; robert.schoelkopf@yale.edu}
\end{affiliations}
\fi

\begin{linenumbers}

\begin{abstract}
    To solve classically hard problems, quantum computers need to be resilient to the influence of noise and decoherence. 
    In such a fault-tolerant quantum computer, noise-induced errors must be detected and corrected in real-time to prevent them from propagating between components \cite{Preskill1998Fault-TolerantComputation,Campbell2017RoadsComputation}.
    This requirement is especially pertinent while applying quantum gates, when the interaction between components can cause errors to quickly spread throughout the system.
    However, the large overhead involved in most fault-tolerant architectures \cite{Fowler2012SurfaceComputation,Campbell2017RoadsComputation} makes implementing these systems a daunting task, which motivates the search for hardware-efficient alternatives \cite{Mirrahimi2014DynamicallyComputation,Guillaud2019RepetitionOverhead}.
    Here, we present a gate enacted by a multilevel ancilla transmon on a cavity-encoded logical qubit that is fault-tolerant with respect to decoherence in both the ancilla and the encoded qubit.
    We maintain the purity of the encoded qubit in the presence of ancilla errors by detecting those errors in real-time, and applying the appropriate corrections. We show a reduction of the logical gate error by a factor of two in the presence of naturally occurring decoherence, and demonstrate resilience against ancilla bit-flips and phase-flips by observing a sixfold suppression of the gate error with increased energy relaxation, and a fourfold suppression with increased dephasing noise.
    The results demonstrate that bosonic logical qubits can be controlled by error-prone ancilla qubits without inheriting the ancilla's inferior performance. As such, error-corrected ancilla-enabled gates are an important step towards fully fault-tolerant processing of bosonic qubits.
\end{abstract}

\ifarxiv\maketitle\fi

In recent years, quantum error correction (QEC) has been demonstrated to protect stored logical qubits against decoherence, either by encoding the information redundantly in a block of multiple physical qubits \cite{Kelly2015StateCircuit,Nigg2014QuantumQubit,Cramer2016RepeatedFeedback,Corcoles2015DemonstrationQubits}, or in a single higher-dimensional bosonic element \cite{Ofek2016ExtendingCircuits,Fluhmann2019EncodingOscillator,Hu2019QuantumQubit}. The concept of fault-tolerant operations extends this principle to the protection of quantum information during a computation involving multiple elements. In particular, errors propagating between elements must not accumulate to the extent that the errors can no longer be removed by QEC. Each task performed in a quantum computer must eventually
be made fault-tolerant, including syndrome measurements \cite{Rosenblum2018Fault-tolerantError.,Linke2017Fault-tolerantDetection}, state preparation \cite{Linke2017Fault-tolerantDetection,Takita2017ExperimentalQubits}, and gates \cite{Harper2019Fault-TolerantExperience}. Remarkably, following this procedure can allow an error-corrected quantum processor to perform computations at any desired accuracy \cite{Aharonov2008Fault-TolerantRate}.

Some quantum gates can be implemented in a way that is naturally protected by the encoding of choice, and do not require additional resources, e.g.\ braiding operations in the surface code \cite{Fowler2012SurfaceComputation}, transversal operations in CSS codes \cite{Steane2005Fault-tolerantCodes}, and displacements in GKP codes \cite{Gottesman2001EncodingOscillator}. However, these ``natural'' operations are often insufficient to create a universal gate set \cite{Eastin2009RestrictionsSets,Webster2018BraidingUniversal}. One method of addressing this shortcoming is to ``inject'' additional gates by coupling to an ancilla qubit that has more complete functionality \cite{Bravyi2005UniversalAncillas,Zhou2000MethodologyConstruction}. These ancilla-based operations are not native to the encoding, and as such, require a significant overhead in hardware and number of operations to implement fault-tolerantly \cite{Fowler2012SurfaceComputation,Campbell2017RoadsComputation}. 

Here, we devise a hardware-efficient circuit that uses a driven ancilla qubit to apply protected gates to a logical qubit encoded in a bosonic mode. Our scheme works by encoding the ancilla in a single multilevel system as well, and using this freedom to identify and correct errors occurring during the gate operation. Remarkably, by employing the information obtained in the final ancilla measurement, we can recover the logical qubit, and reapply the gate if necessary.

The principal mechanism underlying the gate's fault-tolerance is \textit{path-independence} \cite{Ma2019GeneralNoise}---the property that, given fixed initial and final ancilla states, the net logical operation is independent of the specific ancilla trajectory induced by control drives and decoherence events (Figure 1). Path-independence requires that we drive the ancilla in such a way that its populations do not depend on the state of the logical system.
In order to ensure that the evolution of the logical qubit under the effect of decoherence is a fixed unitary, our circuit uses an error-transparent interaction  \cite{Kapit2018Error-TransparentArchitectures,Rosenblum2018Fault-tolerantError.} to couple the ancilla and qubit.
\end{linenumbers}
\newcommand{\figone}{
\begin{figure}
    \centering
    \includegraphics{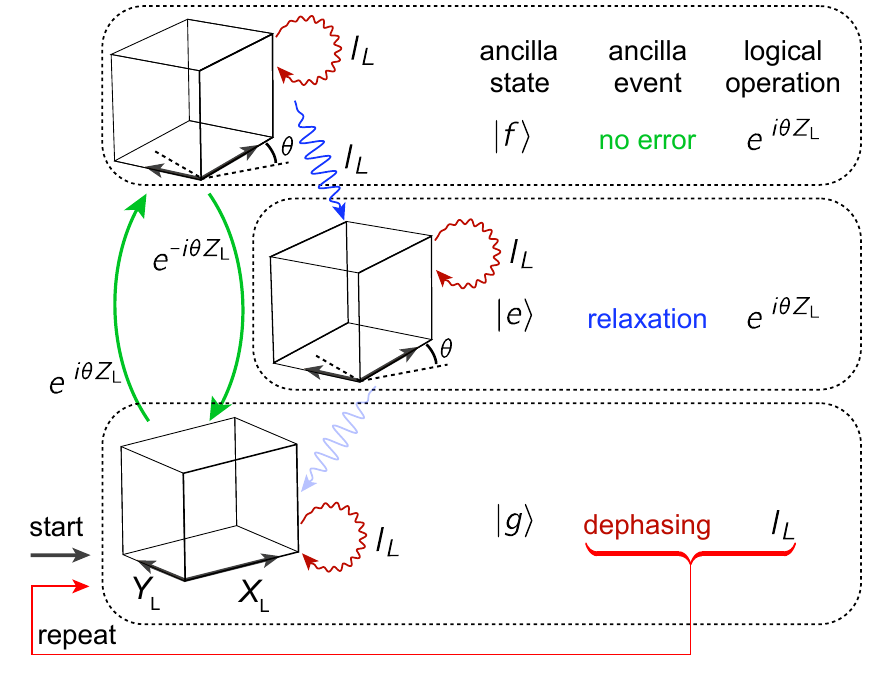}
    \caption{\textbf{Working principle of the error-corrected logical gate}. Control drives excite the ancilla from the ground state $\ket{g}$ to the second excited state $\ket{f}$ around an axis $e^{i\phi}\dyad{f}{g} + e^{-i\phi}\dyad{g}{f}$ with $\phi=\theta$ for  the logical state $\ket{1}_\text{L}$ and $\phi=0$ for $\ket{0}_\text{L}$. This effects a $Z_\text{L}$ rotation (green arrows) by an angle $\theta$ on the logical system (boxes). Path-independence requires that all closed loops in the ancilla transition graph produce an identity operation on the logical qubit, implying that the logical operation is uniquely determined by a measurement of the ancilla state (Supplementary Information section 2).
    A rapid unconditional $\ket{g}\leftrightarrow\ket{f}$ swap (shown in Fig. 2d) is applied before the measurement to minimize the probability of ancilla relaxation during the measurement.
    Error transparency guarantees that the logical operation associated with the dominant decoherence events (ancilla relaxation and dephasing, depicted by blue and dark red arrows) is the identity $I_\text{L}$. The operation succeeds in the case of either no error or relaxation. In the case of dephasing the operation is not applied, but repeating the protocol makes the gate succeed deterministically. Relaxation from $\ket{e}$ to $\ket{g}$, shown with a transparent arrow, breaks path-independence, as well as error transparency, but is a low-probability second-order error.}
    \label{fig1}
\end{figure}
}
\ifarxiv\figone\fi
\begin{linenumbers}

We implement our logical qubit using a single bosonic mode in a superconducting cavity (Fig.\ 2a, Supplementary Information section 8). Within this mode, quantum information is stored in the subspace defined by the binomial ``kitten'' code \cite{Michael2016NewMode}, which, like the Schr\"odinger cat code \cite{Mirrahimi2014DynamicallyComputation}, can correct for the loss of a single photon. The logical states can be represented in the photon number basis as $\ket{0}_\text{L} = \frac{1}{\sqrt{2}}\left(\ket{0} + \ket{4}\right)$ and $\ket{1}_\text{L} = \ket{2}$.
We implement a family of operations $S(\theta)=e^{i\theta Z_\text{L}}$, which are rotations by any angle $\theta$ around the $Z_\text{L}$ axis in the logical subspace (Fig.\ 2b), using the selective number-dependent arbitrary phase (SNAP) protocol \cite{Krastanov2015UniversalQubit,Heeres2015CavityGates}. This protocol drives a dispersively coupled transmon ancilla qubit to the excited state with a photon-number dependent phase. $S(\theta)$ can be effected by choosing the phase of the control drive to be zero on photon number states $\ket{0}$ and $\ket{4}$, and $\theta$ on $\ket{2}$. The arbitrary angle of rotation in $S(\theta)$ allows the realization of logical Clifford operations ($\theta= k\pi/2$, with $k \in \mathbb{Z}$) as well as non-Clifford operations such as the $T$-gate ($\theta=\pi/4$). This gate set can be combined with a single rotation around a different logical axis to provide universal control of the logical qubit.

\end{linenumbers}
\newcommand{\figtwo}{
\begin{figure*}[tb]
    \centering
    \includegraphics{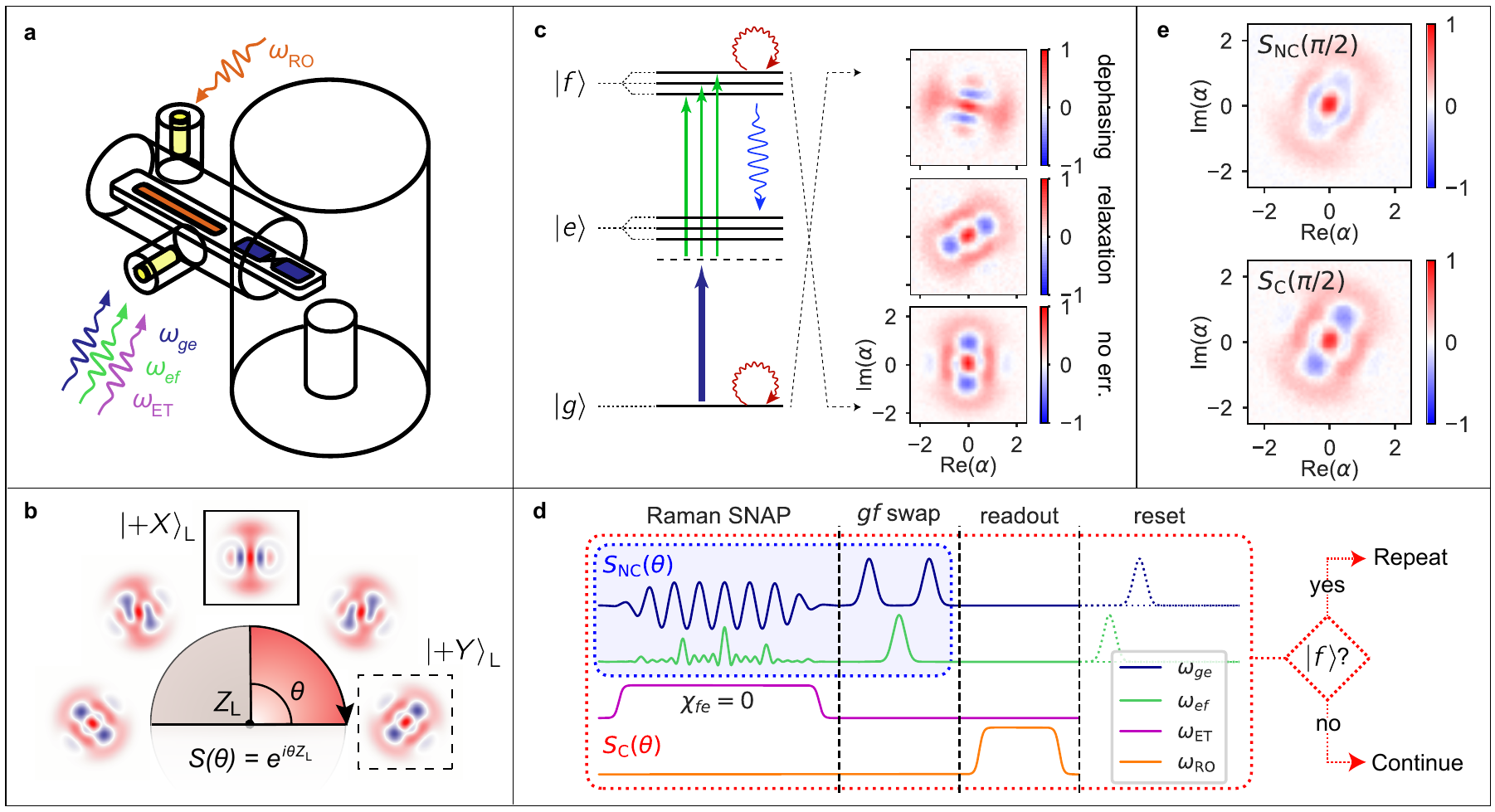}
    \caption{\textbf{Experimental protocol and tomography of logical states after gate application.} \textbf{a}, The system consists of a $\lambda/4$ coaxial superconducting cavity coupled to an ancilla transmon, which is in turn coupled to a stripline readout resonator. The protocol involves control drives, which address the first ($\omega_{ge}$) and second ($\omega_{e\f}$) ancilla transition frequencies off-resonantly, the error-transparency drive (at $\omega_\text{ET}$, Supplementary Information section 6) \cite{Rosenblum2018Fault-tolerantError.} and the readout drive ($\omega_\text{RO}$).
    \textbf{b}, The protocol effects a rotation around the logical $Z_\text{L}$ axis by an arbitrary amount $\theta$. In the following demonstrations, we use the initial state $\ket{+X}_\text{L}$ (simulated Wigner function in the solid box), and a rotation angle of $\theta=\pi/2$, producing the target state $\ket{+Y}_\text{L}$ (dashed box).
    \textbf{c},  The Raman SNAP operation consists of applying a control drive detuned from the $\omega_{ge}$ transition (blue arrow) as well as a comb of control drives (green arrows), detuned in the opposite sense from the $\omega_{e\f}$ transition and separated in frequency by twice the ancilla-cavity dispersive shift $2\chi_{\f g}$ (Supplementary Information section 5). The measured Wigner tomogram of the cavity state, postselected on the final ancilla state following a $g\f$ swap (dashed arrows), is shown on the right.
    \textbf{d}, The control sequence without error correction ($S_\text{NC}(\theta)$, blue box) involves the Raman SNAP operation and a $g\f$ swap. In the error-corrected protocol ($S_\text{C}(\theta)$, red box), we add the error-transparency drive during the SNAP operation in order to remove the random cavity phase space rotation induced by $\ket{f}\rightarrow\ket{e}$ relaxation. We also add an ancilla readout and reset. Additionally, in the error-corrected protocol, we re-apply all steps upon measuring $\ket{f}$.
    \textbf{e}, Unconditional Wigner tomogram after application of the logical gate without error correction (top) and with error correction (bottom). The data in this figure were obtained with artificially induced ancilla dephasing and relaxation probabilities ($\sim$20\% each) to emphasize the increase in the fidelity of the final state when error correction is performed.
    }
    \label{fig2}
\end{figure*}}
\ifarxiv\figtwo\fi
\begin{linenumbers}

The ancilla undergoes two predominant types of errors: dephasing and energy relaxation. To correct the effects of these errors, we must first make them independently detectable. We do so by modifying the SNAP protocol to use three levels of the ancilla, instead of two. By driving a Raman transition (Fig.~2c, Supplementary Information section 5) from the ground state ($\ket{g}$) to the second excited state ($\ket{f}$) with a photon-number dependent phase, we implement the SNAP operation while avoiding population of the first excited state ($\ket{e}$). Next, we swap $\ket{f}$ and $\ket{g}$, so as to minimize the probability of ancilla relaxation during the subsequent ancilla state measurement (Fig.~2d). The measurement outcome determines which (if any) type of ancilla error occurred, as well as the operation effected on the cavity state.

We ensure protection against ancilla dephasing during the SNAP operation by simultaneously driving the ancilla to $\ket{f}$ with equal rates $\Omega$ for all photon number states in the logical subspace. Since the control drives have photon number dependent phases, the ancilla becomes entangled with the logical system. However, the ancilla \textit{population} remains uncorrelated with the logical state during the driven evolution.
Therefore, projecting the ancilla to $\ket{g}$ or $\ket{f}$ at any time during the protocol, as the environment does in the case of dephasing, does not impart any back-action on the logical state.
However, dephasing events will create some probability of not successfully finishing the transit from $\ket{g}$ to $\ket{f}$. 
By considering the effective Hamiltonian in the interaction picture during the operation (Supplementary Information section 1)
\begin{equation}
    H_\text{int} =  \Omega\,S(\theta)\otimes\dyad{f}{g} + \Omega^*\,S(-\theta)\otimes\dyad{g}{f},
    \label{Hint}
\end{equation}
we can see that the logical action associated with going from $\ket{g}$ to $\ket{f}$ is the desired gate $S(\theta)$, whereas the logical action of going from $\ket{f}$ to $\ket{g}$ is the inverse operation $S(-\theta)$. As a result, if the ancilla trajectory ends in $\ket{g}$ ($\ket{f}$ following the final swap) due to a dephasing event, the net effect on the logical system is the identity operation.
Remarkably, this path-independence ensures protection even against multiple dephasing events.
We can ensure deterministic application of the gate in the presence of dephasing by resetting the ancilla and repeating the protocol upon measuring $\ket{f}$.

Energy relaxation during the application of the gate occurs predominantly through decay from $\ket{f}$ to $\ket{e}$. The latter state remains unaffected under the action of the control drives, and therefore the final state should be detected as $\ket{e}$, assuming no further decay events. Since the trajectory taking the ancilla from $\ket{g}$ to $\ket{e}$ passes through $\ket{f}$ (Fig.\ 1), the effective operation on the logical system is $S(\theta)$. However, the cavity state will also acquire a random phase space rotation (depending on the jump time) due to the static cavity-ancilla interaction $\chi_{\f e}\dyad{f}{f}a^\dagger a$, with $a^\dagger a$ the photon-number operator and $\chi_{\f e}$ the dispersive interaction rate in $\ket{f}$ in a frame rotating with $\ket{e}$. This random rotation can be understood as the back-action induced by the emitted ancilla excitation carrying photon-number dependent energy. By using the detuned sideband driving scheme presented in Ref. \subcite{Rosenblum2018Fault-tolerantError.}, we can effectively set $\chi_{\f e} = 0$ for the duration of the gate (Supplementary Information section 6). This ``error-transparency'' drive eliminates the random rotation imparted on the cavity state, thereby maintaining path-independence in the case of relaxation.

In addition to ancilla errors, the protocol is compatible with QEC protecting against photon loss in the cavity. Since the control drives do not act on the system in the odd photon number subspace, the result is equivalent to incomplete driving followed by photon loss. While not done in this work, applying a parity measurement \cite{Sun2014TrackingMeasurements,Rosenblum2018Fault-tolerantError.} and recovery operation \cite{Michael2016NewMode} following the protocol, would make the effect of photon loss equivalent to that of ancilla dephasing (Supplementary Information section 3).

The key feature is that, regardless of the measured ancilla state, the cavity remains in a definite pure state contained within the logical subspace. To demonstrate this, we create \cite{Heeres2017ImplementingOscillator} the state $\ket{+X}_\text{L} = \frac{1}{\sqrt{2}}(\ket{0}_\text{L} + \ket{1}_\text{L})$, apply the operation $S(\pi/2)$, and perform Wigner tomography (Fig.\ 2c). In this experiment, the error-transparency drive is applied, and the ancilla is measured without conditional repetition of the gate. The Wigner functions are shown separately for each measured ancilla state. In order to emphasize the effect of ancilla errors, we increase the ancilla error probability during this operation (Supplementary Information section 4), so that the probability of relaxation and dephasing errors are $\sim\!20\%$ each. As expected, in the case of successful completion of the protocol (ancilla in $\ket{g}$), or in the case of relaxation (ancilla in $\ket{e}$), the gate is correctly applied. Different deterministic phase space rotations are acquired by the cavity for different final ancilla states, as a result of evolution during the ancilla measurement. This angle is corrected in software by updating the phase of subsequent drives around the cavity resonance frequency. Finally, in the case of ancilla dephasing (ancilla in $\ket{f}$), we observe the initial logical state $|+X\rangle_\textrm{L}$. The slight asymmetry is a result of the Kerr evolution, whose removal requires successful completion of the logical gate \cite{Heeres2015CavityGates}.

We next perform two versions of the full gate, the standard (non-corrected) gate $S_\text{NC}$ and the error-corrected gate $S_\text{C}$ (Fig.\ 2d), again with increased ancilla error rates. We characterize the result via Wigner tomography without conditioning on the ancilla measurement outcome (Fig.\ 2e). In the case of the standard gate ($S_\text{NC}$), we observe significant smearing of the final state.
However, in the case of the error-corrected gate ($S_\text{C}$), which includes the error-transparency drive, ancilla measurement, reset, and conditional repetition, it is evident that, despite the high ancilla error rate, the cavity coherence is mostly preserved.

In order to establish the gate's logical error probability quantitatively, we turn to interleaved randomized benchmarking (IRB) \cite{Magesan2012EfficientBenchmarking}. We first create a set of operations from the logical Clifford group using numerical optimal control \cite{Heeres2017ImplementingOscillator}. We then interleave the $S(\pi/2)$ gate between randomly selected Clifford operations, scanning the length of the sequence (Fig.\ 3a). We measure the probability of obtaining the correct answer in the ancilla after applying a decoding operation as a function of the sequence length $n$ (Fig.\ 3b), and compare the performance of $S_\text{NC}$ and $S_\text{C}$. The measured gate error probability for $S_\text{NC}$ (obtained from the difference between the decay rates of the interleaved and the non-interleaved sequences) is $4.6\%\pm 0.1\%$. The main effect in producing this error probability is ancilla relaxation ($\sim 2.5\%)$, with an additional $0.8\%$ accounted for resulting from ancilla dephasing, photon loss, and thermal ancilla excitation. In contrast, the error-corrected gate has an error probability of $2.4\%\pm 0.1\%$. No single process dominates the remaining error, but a full accounting of the known sources, including photon loss, readout-induced dephasing, and other sources, predicts an error probability of 2.1\% (Supplementary Information section 7). 

\end{linenumbers}
\newcommand{\figthree}{
\begin{figure*}[tb]
    \centering
    \includegraphics{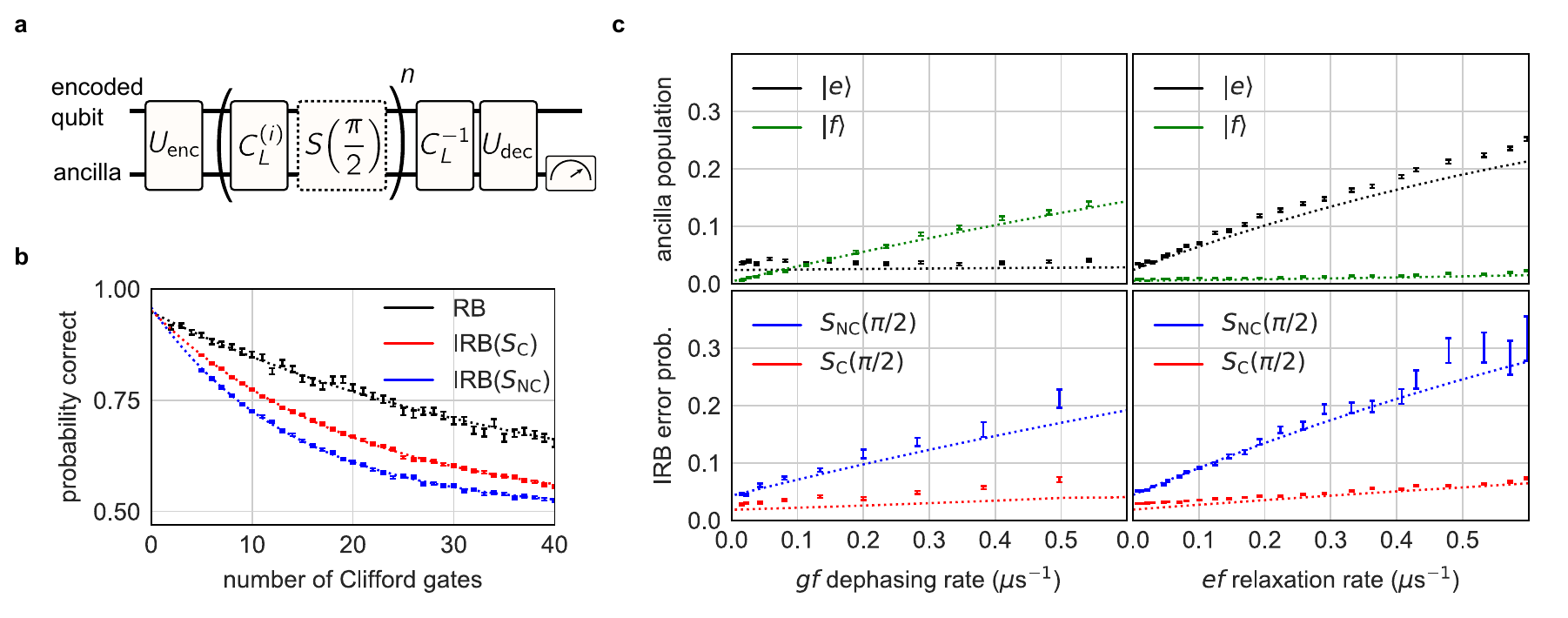}
    \caption{\textbf{Benchmarking of the logical gate.} \textbf{a}, Circuit for randomized benchmarking (RB) and interleaved randomized benchmarking (IRB) \cite{Heeres2017ImplementingOscillator}. We first prepare an encoded state of the cavity using a numerically optimized encoding pulse $U_\text{enc}$. Next, we perform a random sequence of $n$ operations drawn from the Clifford group, $C_\text{L}^{(i)}$. In the interleaved variants, after each random Clifford, we apply the logical gate (dashed box), either using the non-error-corrected ($S_\text{NC}$) or error-corrected ($S_\text{C}$) protocols. Finally, the net inverse Clifford operation is applied, followed by a decoding operation $U_\text{dec}$. This maps the encoded information onto the ancilla, where it can be measured. \textbf{b}, By fitting both the RB and IRB results to an exponential model $A e^{-\gamma n}+\frac{1}{2}$ (dotted lines), we can learn the effective gate error probability. Without interleaved logical gates, we measure $\gamma_\text{RB} = 2.5\% \pm 0.1\%$ (black). We can determine the error probability associated with the error-corrected (non-error-corrected) operation as $\gamma_\text{IRB} - \gamma_\text{RB} = 2.4\% \pm 0.1\%$ ($4.6\% \pm 0.1\%$) from the red (blue) curve. \textbf{c}, In order to demonstrate the robustness of the protocol, we add noise to the system (Supplementary Information section 4), which has the effect of increasing either the ancilla dephasing (left panels) or relaxation rates (right panels). We measure the ancilla population (top panels) and IRB-inferred error probabilities (bottom panels). We see that the populations are affected nearly independently by the respective noise parameters. We also see that in both cases, $S_\text{C}$ (red markers) is significantly less likely than $S_\text{NC}$ (blue markers) to translate ancilla errors induced by the added noise into logical errors. The dotted lines are derived from a full quantum simulation using independently measured system parameters.}
    \label{fig3}
\end{figure*}}
\ifarxiv\figthree\fi
\begin{linenumbers}

In order to demonstrate the robustness of the gate to ancilla decoherence, we intentionally introduce noise to increase the ancilla dephasing and relaxation rates (Supplementary Information section 4). We scan the induced $g\f$ dephasing rate and $e\f$ relaxation rates from their native rates of $1/(40\,\mu\text{s})$ and $1/(47\,\mu\text{s})$, respectively, up to maximum rates of $\sim 1/(2\,\mu\text{s})$. The ancilla state probabilities vary as expected, with $P_e$ changing from 3.4\% to 25\% with increased relaxation rate and $P_f$ changing from $<$1\% to 14\% with increased dephasing rate (Fig.\ 3c). 
For $S_\text{NC}$, the induced gate error probability is nearly equal to the probability of measuring the ancilla in an excited state, indicating that ancilla errors are bound to propagate and affect the logical qubit. However, for $S_\text{C}$, the ratio of gate errors to ancilla errors (Supplementary Information section 4) is suppressed by a factor of $5.8 \pm 0.2$ ($4.2 \pm 0.4$) for injected $e\f$ relaxation ($g\f$ dephasing) errors, clearly demonstrating the resilience of the gate against ancilla errors.  


We have presented an ancilla-enabled gate that uses the principle of path-independence to maintain coherence of the logical qubit in the presence of ancilla errors. We demonstrated that the error-correction suppresses the propagation of the dominant errors, namely relaxation and dephasing, from the ancilla to the logical system. The scheme is readily extended to protect against a broader class of ancilla errors, such as thermal excitation or multiple-decay events, by employing higher ancilla levels. Furthermore, we can incorporate protection against photon loss by performing a fault-tolerant parity measurement \cite{Rosenblum2018Fault-tolerantError.} after the gate, and using the result to perform QEC \cite{Ofek2016ExtendingCircuits}. We have demonstrated the feasibility of a hardware-efficient approach to protecting quantum information not only during storage, but also as it is being processed by quantum gates. Expanding these results to create additional error-corrected gates \cite{Xu2018GeometricallyGates,Gao2019EntanglementInteraction}, and therefore providing universal fault-tolerant control, is a promising path towards robust quantum computing devices.

\end{linenumbers}
\ifarxiv
\bibliography{main}
\else
\clearpage
\printbibliography
\fi

\newcommand{\theacknowledgements}{We thank N. Frattini and K. Sliwa for providing the Josephson parametric converter and N. Ofek for providing the logic for the field programmable gate array (FPGA) used for the control of this experiment. We thank M. Zhang and Y. Wang for helpful discussions. S.R., L.F. and R.J.S. acknowledge funding support from the U.S. Army Research Office (W911NF-18-1-0212). P.R. and S.R. were supported by the Air Force Office of Scientific Research (FA9550-15-1-0015 and FA9550-14-1-0052).

}
\newcommand{\authorcontributions}{ P.R. and S.R. fabricated the transmon qubits, assembled the experimental apparatus, performed the experiments, and analyzed the data under the supervision of L.F. and R.J.S. W.M. and L.J. provided theoretical support. P.R., S.R., and R.J.S. wrote the manuscript with feedback from all authors.}
\newcommand{\data}{The data that support the findings of this study are available from the corresponding authors upon reasonable request.}
\newcommand{\additionalinfo}{Reprints and permissions information is available at www.nature.com/reprints. Correspondence and requests for materials should be addressed to P.R., S.R. or R.J.S.}
\newcommand{\competing}{L.F. and R.J.S. are co-founders of, and equity shareholders in, Quantum Circuits, Inc. S.R., P.R., L.J., L.F. and R.J.S. are inventors on patent application no. 62/613,974 submitted by Yale University, which covers hardware-efficient fault-tolerant operations with superconducting circuits.}
\newcommand{\supp}{is linked to the online version of the paper at www.nature.com/nature.}

\ifarxiv

\begin{acknowledgements}
\begin{center}
\textbf{Acknowledgements\\}
\end{center}
\theacknowledgements
\end{acknowledgements}

\else
\begin{addendum}
 \item[Supplementary Information] \supp 
 \item \theacknowledgements
 \item[Data availability] 
 \data 
 \item[Author contributions] \authorcontributions
  \item[Competing interests] \competing
 \item[Additional information] \additionalinfo
\end{addendum}
\fi

\clearpage

\ifarxiv\else
\singlespacing
\figone
\figtwo
\figthree
\fi

\clearpage

\ifarxiv
\onecolumngrid
\fi


\setcounter{equation}{0}
\setcounter{figure}{0}
\setcounter{table}{0}
\setcounter{page}{1}
\makeatletter
\renewcommand{\thesection}{S\arabic{section}}
\renewcommand{\theequation}{S\arabic{equation}}
\renewcommand{\thefigure}{S\arabic{figure}}
\renewcommand{\thetable}{S\Roman{table}}
\newcommand{\opH}{H}
\newcommand{\opa}{a}
\newcommand{\opad}{a^\dagger}
\newcommand{\opsig}{\sigma}
\newcommand{\hc}{\text{h.c.}}
\newcommand{\Hint}{\opH_\text{int}}
\newcommand{\proj}[1]{\dyad{#1}{#1}}
\newcommand{\opU}{U}
\newcommand{\unit}{\ \mathrm}
\newcommand{\opb}{b}
\newcommand{\opbd}{b^\dagger}

\begin{center}
\textbf{\Large Supplementary Information for: \\ Error-corrected gates on an encoded qubit}
\end{center}
\author{Philip Reinhold$^{1,2}*\dag$, Serge Rosenblum$^{1,2,3}*\dag$, Wen-Long Ma$^{1,2}$, Luigi Frunzio$^{1,2}$, Liang Jiang$^{1,2,4}$ \& Robert J. Schoelkopf$^{1,2}*$}

\ifarxiv\else
\begin{affiliations}
\blfootnote{
 \item Department of Applied Physics and Physics, Yale University, New Haven, CT, USA
 \item Yale Quantum Institute, Yale University, New Haven, CT, USA
 \item Current address: Department of Condensed Matter Physics, Weizmann Institute of Science, Rehovot, Israel
 \item Current address: Pritzker School of Molecular Engineering, University of Chicago, Chicago, Illinois 60637, USA
 \item[] $\dag$ These authors contributed equally
 \item[] * e-mail: philip.reinhold@yale.edu; serge.rosenblum@weizmann.ac.il robert.schoelkopf@yale.edu
}
\end{affiliations}
\begin{refsection} 
\fi

\begin{linenumbers}
\ifarxiv\else\resetlinenumber\fi
\section{An interaction picture for SNAP}
The SNAP operation \subcite{Krastanov2015UniversalQubit} consists of several control drives simultaneously applied to a transmon ancilla, which is dispersively coupled to a superconducting cavity. The control drive frequencies are detuned from vacuum resonance by an integer multiple of the dispersive shift $\chi$. In this paper, we address a logical state stored within the even photon number parity subspace of the cavity mode, and therefore, drive the ancilla only at frequencies detuned by multiples of $2\chi$. Each of these drives has the same driving rate $\Omega$, but differs in the phase $\theta_k$. We can write the interaction Hamiltonian as follows:
\begin{equation}
    \opH = \frac{\chi}{2}\opad\opa\opsig_z + \Omega \sum_{k\,\text{even}} e^{i(k\chi t + \theta_k)}\opsig_- + \hc
\end{equation}
We can modify this Hamiltonian by considering transitions directly between $\ket{g}$ and the third ancilla level $\ket{f}$.
\begin{equation}\label{eq:ft-snap-H-int}
    H = \left(\chi_e\proj{e}  + \chi_f\proj{f}\right)\opad\opa
    + \Omega \sum_{k\,\text{even}} e^{i(\chi_f k t + \theta_k)}\dyad{g}{f} + \hc
\end{equation}
In order to obtain a time-independent picture for this operation, we perform the canonical transformation using the time-dependent unitary
\begin{equation}
    \opU = \exp{it\left(\chi_e\proj{e} + \chi_f\proj{f}\right)\opad\opa}.
\end{equation}
Under this transformation, the ladder operators are transformed as follows:
\begin{align}
    \opa &\mapsto e^{i(\chi_e\proj{e} + \chi_f\proj{f})t}\opa \label{eq:snap-frame-opa}\\
    \dyad{g}{f} &\mapsto e^{i\chi_f t\opad\opa}\dyad{g}{f} \label{eq:snap-frame-gf}\\
    \dyad{g}{e} &\mapsto e^{i\chi_e t\opad\opa}\dyad{g}{e} \label{eq:snap-frame-ge}\\
    \dyad{e}{f} &\mapsto e^{i(\chi_f-\chi_e) t\opad\opa}\dyad{e}{f}. \label{eq:snap-frame-ef}
\end{align}
The jump operators are also transformed in precisely this way. The resulting interaction picture Hamiltonian can be written as follows:
\begin{align}
    \Hint &= \Omega \sum_{k\,\text{even}} e^{i(\chi_f k t - \chi_f\opad\opa t + \theta_k)}\dyad{f}{g} + \hc\\
    &\approx \Omega \sum_{k\,\text{even}} e^{i\theta_k}\dyad{f,k}{g,k} + \hc,\label{eq:snap-h-int-approx}
\end{align}
where the rotating wave approximation that we have made in the second line is valid in the limit where $\chi_f \gg \Omega$. We can simplify this expression by defining the effective unitary operation on the even photon number subspace effected by the SNAP operation
\newcommand{\snap}{S(\vec{\theta})}
\newcommand{\msnap}{S(-\vec{\theta})}
\begin{equation}
    \snap = \sum_{k\,\text{even}} e^{i\theta_k}\proj{k}.
\end{equation}
We can therefore rewrite \ref{eq:snap-h-int-approx} as:
\begin{align}
    \Hint = \Omega\left(\snap \dyad{f}{g} + \msnap\dyad{g}{f}\right).
\end{align}
This Hamiltonian is ``Pauli-like,'' in that it squares to the identity within the driven subspace:
\begin{align}
    \Hint^2  &= \Omega^2 \sum_{k\,\text{even}} \proj{g,k} + \proj{f,k}\\
    &\equiv \Omega^2 P,\label{eq:driven-subspace}
\end{align}
where $P$ is the projector on the driven subspace. Therefore, we obtain the propagator
\begin{equation}
    e^{-i\Hint t} = (\mathbb{I} - P) + \cos(\Omega t)P - i\sin(\Omega t)\left(\snap\dyad{f}{g} + \msnap\dyad{g}{f}\right).\label{snap-evolution}
\end{equation}
In practice, $\Omega$ will not be constant, but rather have some Gaussian envelope profile. In this case, we can replace $\Omega t$ with the integrated area under the envelope.

\section{Error-transparency and path-independence}
\label{sec:ft-snap:error-prop}
We now describe the properties of error-transparency and path-independence in a general setting. We consider a Hamiltonian that can be split into two terms, the static interaction Hamiltonian $H_0$, and the time-dependent controls $H_c(t)$:
\begin{equation}
    H = H_0 + H_c(t).
\end{equation}
In addition, there are a set of jump operators $\{J_k\}$ used in the Lindblad master equation:
\begin{equation}
    \partial_t \rho = \frac{i}{\hbar}\comm{H}{\rho} + \sum_k J_k\rho J_k^\dagger - \frac{1}{2}(J_k^\dagger J_k\rho + \rho J_k^\dagger J_k).
\end{equation}
The \textit{error-transparency} condition is satisfied when the jump operators commute (up to a dephasing operation on the ancilla) with the evolution generated by $H_0$: \footnote{More rigorously, we should also include the non-Hermitian part of the Hamiltonian (associated with the last term in Eq. S16). For ancilla relaxation and dephasing jump operators, however, the non-Hermitian part of the Hamiltonian actually satisfies the error transparency condition. Hence, for simplicity, we will neglect them in our derivation below.}
\begin{equation}
    \forall_k:\,\comm{H_0}{J_k} = J_k H_A \Longrightarrow e^{iH_0 (T-t)} J_k e^{iH_0 t} = J_k e^{iH_A (T-t)} e^{iH_0 T},
\end{equation}
where $H_A$ is an operator acting on the ancilla, for which $[H_0,H_A]=0$.
In our case $H_0$ is given by the dispersive interaction:
\begin{equation}
    H_0 = \opad\opa(\chi_e\proj{e} + \chi_f\proj{f}),
\end{equation}
and the jump operators correspond to photon loss from the cavity $a$, ancilla dephasing $\proj{f}$, or ancilla relaxation $\dyad{e}{f}$. Examining the corresponding error-transparency conditions:
\begin{align}
    \comm{H_0}a &= a(\chi_e\proj{e} + \chi_f\proj{f})\label{loss}\\    \comm{H_0}{\proj{f}} &= 0\\
    \comm{H_0}{\dyad{e}{f}} &= (\chi_e - \chi_f)\opad\opa\dyad{e}{f},
\end{align}
we see that we can satisfy these by setting $\chi_e = \chi_f$, which we achieve using the detuned sideband drive approach detailed in Ref. \subcite{Rosenblum2018Fault-tolerantError.}.

Stating the \textit{path-independence} criterion in a mathematical formalism requires more effort that will be detailed in Ref. \subcite{Ma2019GeneralNoise}, but an intuitive picture can be described succinctly. We note that the ancilla jump operators induce a preferred basis for the ancilla $\{\ket{n}\}$. We can represent the Hamiltonian, in the interaction picture $H_\text{int} = e^{iH_0t}H_ce^{-iH_0t}$, in the following way:
\begin{equation}
    H_\text{int} = \sum_{n,m} \dyad{n}{m}\otimes M_{n,m},
\end{equation}
where $M_{n,m}$ is an operator acting on the logical subsystem \subcite{Ma2019GeneralNoise}. We can construct a graph (Fig.\ 1 in the main text) with nodes corresponding to ancilla states $\ket{n}$ and edges corresponding to transitions, either resulting from $H_\text{int}$ or the jump operators. To each edge we can assign an action on the logical system.
As a result of error transparency, edges corresponding to ancilla jump operators have an associated logical action of the identity. The control-associated transitions are given by $M_{n,m}$ where this value is non-zero. To achieve path-independence, we would like the action associated with any closed loop on the graph to be identity. This condition can be satisfied by unitary matrices $M_{n,m}$, such that $M_{n,m} = M_{m,n}^{-1}$. Moreover, we must check that the addition of the jump operator edges adds no loops with non-trivial net logical action. For instance, in the case of the graph shown in Fig.\ 1b, the addition of the $\dyad{e}{f}$ transition adds no loops, but adding the second-order $\dyad{g}{e}$ transition would form a closed loop with the action $e^{i\theta Z_L}$, which violates path-independence.

\section{Analyzing Fault-Propagation during the SNAP operation}

We now elaborate on how path-independence and error-transparency result in fault-tolerance of the SNAP operation against decoherence. We consider what happens if a discrete jump happens at some time $t$ in the middle of the operation whose total time is $T$. We first consider a relaxation event $\dyad{e}{f}$. Recalling equation~\ref{eq:snap-frame-ef}, in the interaction picture we must write $e^{i(\chi_f - \chi_e)t\opad\opa}\dyad{e}{f}$. We assume we start with the ancilla in the ground state, and some state in the cavity $\ket{\psi_\text{cav}}$. We analyze the evolution in three steps: an initial Hamiltonian evolution, the application of a jump operator, and the remaining Hamiltonian evolution. 
\begin{align}
    \ket{\psi_\text{final}} &\propto e^{i\Hint (T - t)} \left(e^{i(\chi_f-\chi_e)t\opad\opa}\dyad{e}{f} \right)e^{i\Hint t} \left(\ket{\psi_\text{cav}}\otimes\ket{g}\right)\\
    &\propto e^{i(\chi_f-\chi_e)t\opad\opa}\left(\snap\ket{\psi_\text{cav}}\right)\otimes\ket{e}.
\end{align}
We see that the logical operation on the cavity state still takes place\footnote{This is only the case when we start in the ground state $\ket{g}$. We can implement the SNAP operation starting in $\ket{f}$, with very similar results, except that in the case of relaxation ($\dyad{e}{f}$), the effective logical operation is the identity, as in the case of ancilla dephasing.}. The intuition here is that, in order to have a relaxation event in the first place, the ancilla must have made a transit from $\ket{g}$ to $\ket{f}$. However, the evolution includes an unwanted rotation $e^{i(\chi_f-\chi_e)t\opad\opa}$, which depends on the random jump time $t$, and is therefore not deterministic. We remove this random rotation via the error-transparency drive \subcite{Rosenblum2018Fault-tolerantError.}, which ensures that $\chi_e = \chi_f$. After a relaxation event the ancilla ends up in the incorrect state $\ket{e}$. This is addressed by measuring and resetting the ancilla as part of our protocol.

In the case of ancilla dephasing, the operator we wish to consider is $\proj{f}$. In this case the rotating frame has no effect on the jump operator\footnote{A similar analysis can be performed for different models of dephasing, say $\proj{g}$ or $\proj{f} - \proj{g}$, and yield equivalent results.}:
\begin{align}
    \ket{\psi_\text{final}} &\propto e^{i\Hint (T - t)} \proj{f} e^{i\Hint t} \left(\ket{\psi_\text{cav}}\otimes\ket{g}\right)\\
    &\propto e^{i\Hint (T - t)} \left(\snap\ket{\psi_\text{cav}}\right)\otimes\ket{f}\\
    &\propto \cos\left((\Omega (T - t)\right) \left(\snap\ket{\psi_\text{cav}}\right)\otimes\ket{f} + i\sin\left(\Omega(T - t)\right)\left(\ket{\psi_\text{cav}}\otimes\ket{g}\right).
\end{align}
As before, the act of measuring the ancilla at the end of the protocol simplifies the considerations. We either measure $\ket{f}$ and obtain $\snap\ket{\psi_\text{cav}}$, the desired final state, or we measure $\ket{g}$ and obtain $\ket{\psi_\text{cav}}$, i.e.\ the original state with no operation performed. This error can be remedied by reapplying the gate.

Finally, we consider cavity decay, $\opa$. As in the case of ancilla relaxation, the action of the jump operator takes us out of the driven subspace $P$ (Eq.\ \ref{eq:driven-subspace}), and the evolution freezes as soon as the jump occurs. We can visualize this using the expanded transition graph in Fig.~\ref{fig:expanded-transition-graph}.

\begin{figure}[tb]
    \centering
    \includegraphics{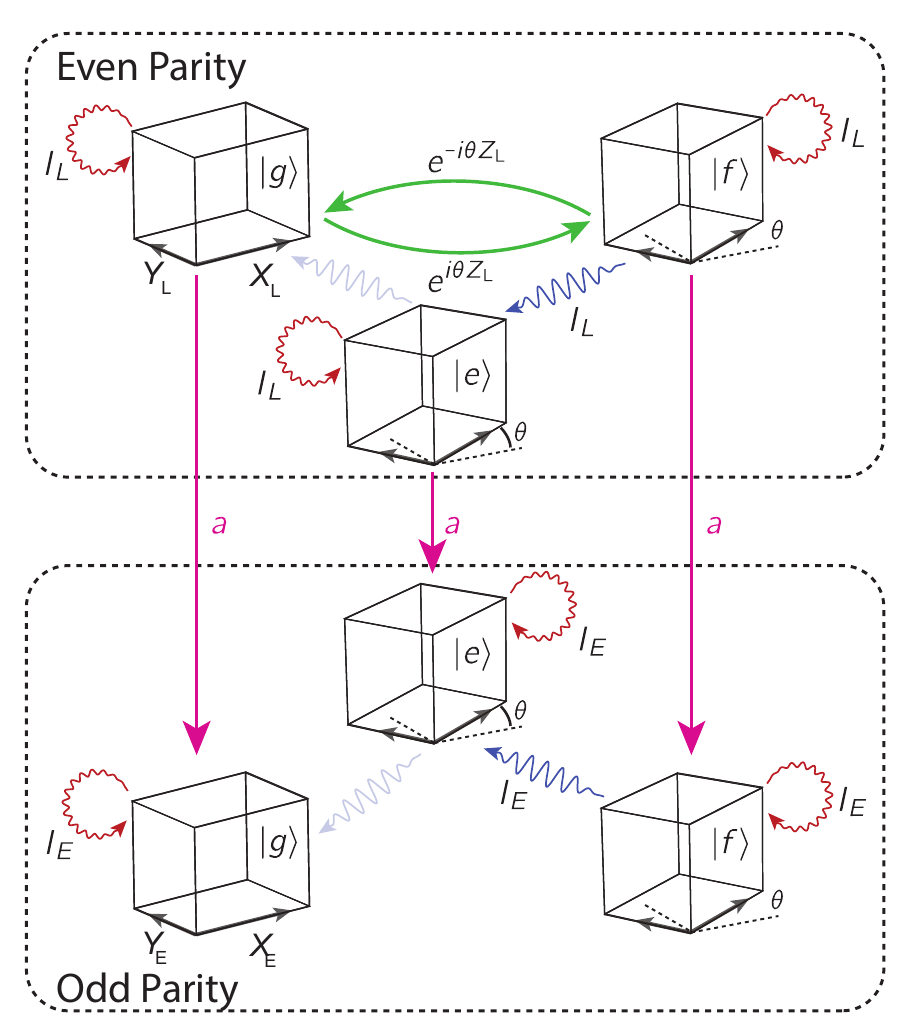}
    \caption{\textbf{Complete transition graph}. In addition to ancilla transitions, we can add transitions in the degree of freedom corresponding to the cavity parity (logical error syndrome).  Logical operators in the odd parity error space are represented by the subscript ``E''. The SNAP pulse has spectral content near the ancilla frequencies associated with photon number states $\ket{0}$, $\ket{2}$ and $\ket{4}$. By making the pulse long, and therefore spectrally narrow, we can reduce the residual spectral content near the ancilla frequencies associated with photon number states $\ket{1}$ and $\ket{3}$. In this case, we can justify omitting arrows between $\ket{g}$ and $\ket{f}$ in the odd-parity subspace, and therefore path-independence is preserved. Relaxation from $\ket{e}$ to $\ket{g}$, shown with a transparent arrow, breaks path-independence, as well as error transparency, but is a low probability second-order error.
}
    \label{fig:expanded-transition-graph}
\end{figure}
Recalling equation~\ref{eq:snap-frame-opa}, $a$ has an $e^{i(\chi_e\proj{e} + \chi_f\proj{f})t}$ time dependency.
\begin{align}
    \ket{\psi_\text{final}} \propto&\, e^{i\Hint (T - t)} \left(e^{i(\chi_e\proj{e}+\chi_f\proj{f})t}\opa\right) e^{i\Hint t} \left(\ket{\psi_\text{cav}}\otimes\ket{g}\right)\\
    \propto&\, \cos(\Omega t) \left(\opa\ket{\psi_\text{cav}}\otimes\ket{g}\right) + 
    i\sin(\Omega t)\left(\opa\snap\ket{\psi_\text{cav}}\otimes\ket{f}\right).
    \label{eq:snap-photon-loss-intermediate}
\end{align}
Therefore, a final measurement of the ancilla will still determine whether the gate was applied. A subsequent parity measurement could detect the photon loss event, which could then be either tracked or corrected, depending on the encoding. In order for this argument to work, we require that the approximation used to produce the effective interaction Hamiltonian (Eq.\ \ref{eq:snap-h-int-approx}) be valid. More specifically, the control drives should leave the odd photon number states unperturbed.

\section{Inducing ancilla decoherence}
In order to demonstrate the robustness of the gate to ancilla decoherence, we employ tools to controllably introduce noise that artificially increases the ancilla dephasing and relaxation rates. To increase the dephasing rate, we add a weak drive tone at the readout mode's resonant frequency $\omega_\text{RO}$, increasing the steady-state population of the readout mode. To increase the relaxation rate, we introduce white noise with 18 MHz bandwidth centered at the $\omega_{e\f}$ transition frequency. Unlike the naturally occurring transitions, the effect of this drive is symmetrical, inducing both $\ket{f}\rightarrow\ket{e}$ as well as $\ket{e}\rightarrow\ket{f}$ transitions in equal measure. At each readout drive ($e\f$ noise) amplitude, we measure the $T_2$ Ramsey ($T_1^{e\f}$) decay curves in order to characterize the $g\f$ dephasing ($e\f$ relaxation) time This allows us to calibrate the $x$-axes of the curves shown in Fig.~3c of the main text.

In order to quantify the susceptibility of the gate to the effect of ancilla decoherence, we can compare the IRB-inferred error probability to the probability of measuring the ancilla in either $\ket{e}$ or $\ket{f}$ (Fig.~\ref{fig:irb_vs_prob}). 

\begin{figure}[htb]
    \centering
    \includegraphics{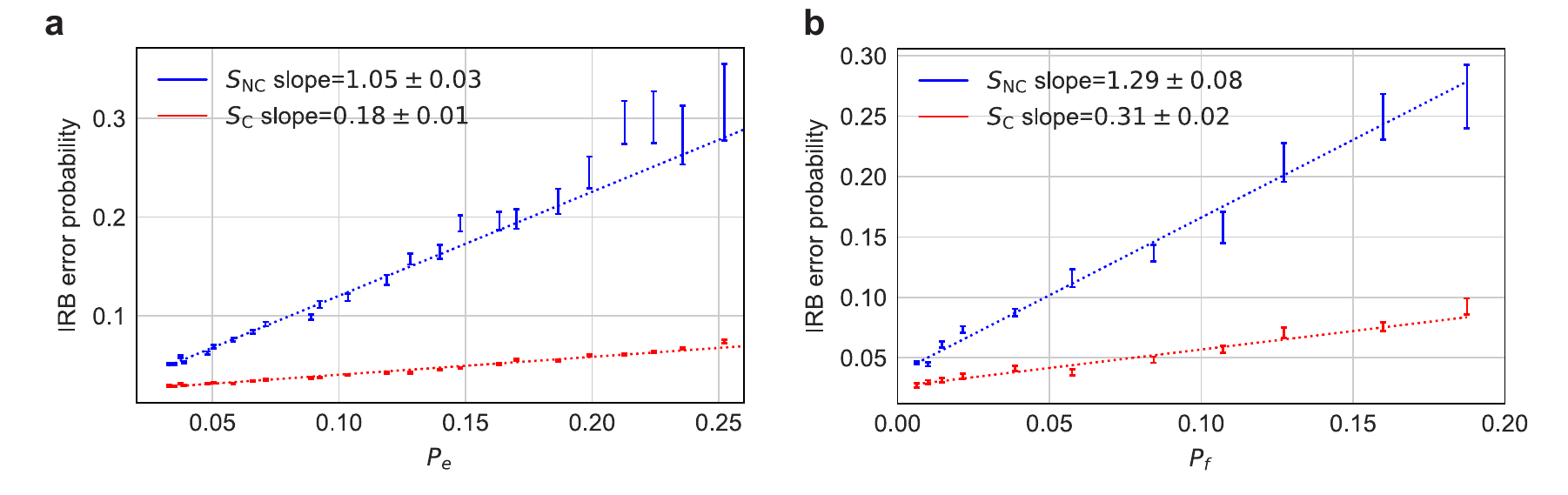}
    \caption{\textbf{Susceptibility of the logical gate to ancilla decoherence.} We measure the IRB-inferred error probabilities vs. the noise-induced probability of an $e\f$ relaxation error (\textbf{a}), and the probability of a $g\f$ dephasing error (\textbf{b}) for $S_\text{NC}$ (blue markers) and $S_\text{C}$ (red markers). The dotted lines are linear fits. The ratio of the slopes for $S_\text{NC}$ and $S_\text{C}$ is used to infer the suppression of gate errors quoted in the main text. The slope of $S_\text{NC}$ is higher than 1 in both cases, since second-order errors (predominantly $\ket{e}\rightarrow\ket{g}$ relaxation and decoherence-associated back-action, respectively) result in a probability of logical dephasing even if the measurement outcome is $\ket{g}$.
    }\label{fig:irb_vs_prob}
\end{figure}

\section{Raman drives for $g\f$ SNAP}
In this section we describe the implementation of a direct $g\f$ drive to implement the $g\f$ SNAP operation described in Eq. 1 of the main text. A single drive, applied at frequency $\omega_{g\f} = \omega_{ge} + \omega_{e\f}$ cannot implement this as a result of the symmetry of the ancilla cosine Hamiltonian (the drive only couples states of differing parity). Instead, we use the method of stimulated Raman transitions, in which we apply control drives to both the $\ket{g}\leftrightarrow\ket{e}$ and $\ket{e}\leftrightarrow\ket{f}$ transitions. We detune these drives by an equal and opposite amount, resulting in frequencies $\omega_{ge} - \Delta$ and $\omega_{e\f} + \Delta$. If $\Delta$ is sufficiently large compared to the drive amplitude, then the effect of this scheme is to drive transitions between $\ket{g}$ and $\ket{f}$ without any intermediate occupation of $\ket{e}$. More accurately, with our parameters the occupation of $\ket{e}$ is limited to approximately $\frac{\Omega_{ge}\Omega_{e\f}}{\Delta^2} \approx 2\%$.  Given individual drive amplitudes $\Omega_{ge}$, $\Omega_{e\f}$, the effective $g\f$ Rabi rate is $\Omega = \frac{\Omega_{ge}\Omega_{e\f}}{\Delta}$.

We wish to combine stimulated Raman driving with the simultaneous number-selective driving of the SNAP operation. In order to do this, we must engineer a situation where, for each $n$ up to the maximum number of addressed photons, there exists a pair of drives with frequencies that satisfy $\omega_1 + \omega_2 = \omega_{g\f} + n\chi_f$. In addition, these frequencies must avoid $\omega_{ge}$ and $\omega_{g\f}$ individually, and we must be careful not to drive spurious transitions by other pairs of drives not considered.

One way of satisfying all of these constraints is to adopt the approach shown in Fig. 2c of the main text. One strong drive is placed at a given detuning $\Delta$ from $\omega_{ge}$ (We choose $\Delta/2\pi =  45\unit{MHz}$). This drive is shared among all photon-number selective transitions. The matching pair for each of these transitions is then given by a weaker tone at $\omega_{e\f} + \Delta + n\chi_f$. In comparison with schemes where every Raman transition has a distinct pair of drive tones, this scheme is much simpler, and avoids the problem of accidentally driving unintended transitions with other tone pairings. 
The ancilla trajectory induced by the Raman SNAP operation is shown in Fig.\ \ref{fig:traj-combined}. An important feature of the trajectory is that the population in $\ket{f}$ shows a $\sim$10\% dependence on the cavity photon number. This violates the condition of path-independence, and will therefore lead to propagation of ancilla errors due to decoherence-induced back-action. The reason for this non-ideal behavior is that the rotating wave approximation in Eq.\ \ref{eq:snap-h-int-approx} is only approximately justified due to the finite ratio of $\Omega/2\chi_f \approx 0.2$. This effect can be mitigated by moving to slower SNAP operation, at the cost of increased decoherence.

\begin{figure}[tb]
    \centering
    \includegraphics[height=.45\textheight]{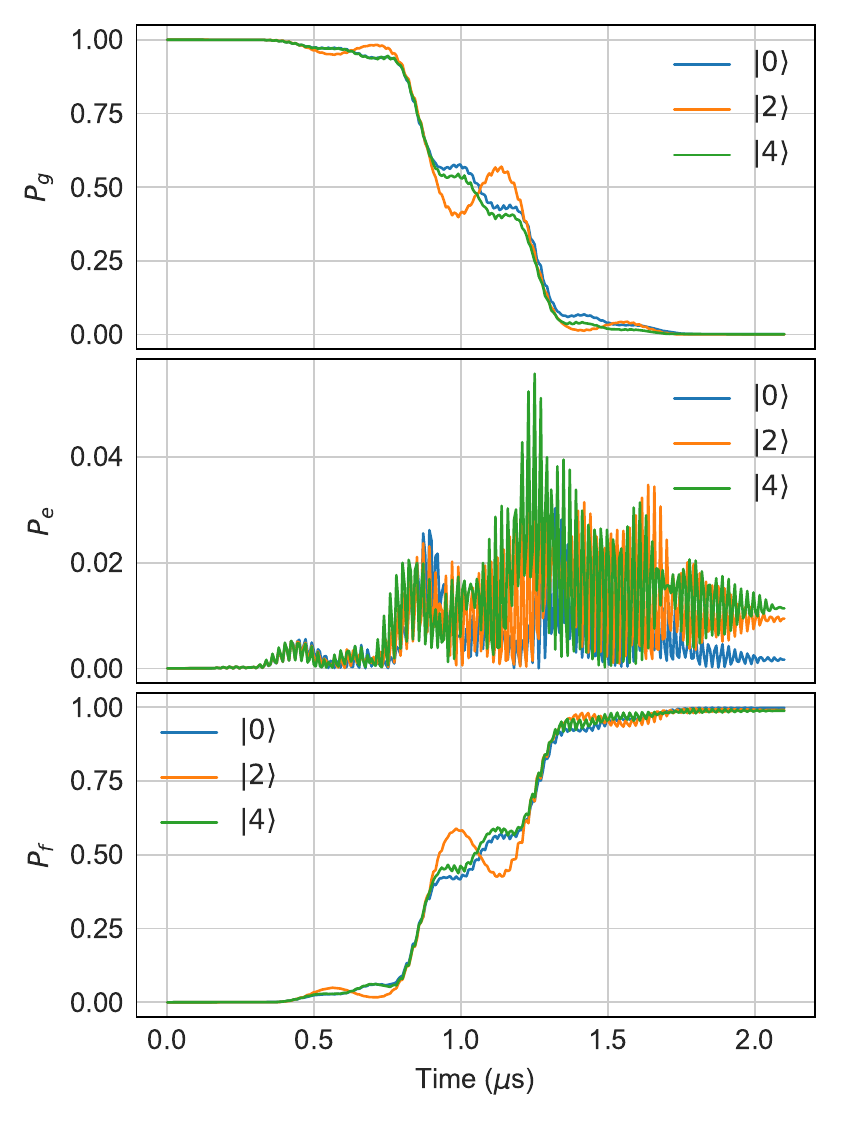}
    \caption{\textbf{Simulated ancilla population trajectory during the Raman SNAP pulse.} The populations in $\ket{g}$, $\ket{e}$, and $\ket{f}$ are shown for $\ket{0}$, $\ket{2}$, and $\ket{4}$ photons in the cavity, which together span the logical subspace. The small difference between the trajectories ($\sim$10\%) causes a small amount of dephasing back-action on the logical system resulting from ancilla decoherence.}
    \label{fig:traj-combined}
\end{figure}

The unconditional $g\f$ swap that is applied before the final ancilla measurement requires a fast transition, and is therefore implemented with direct $ge$ and $e\f$ drives (Fig. 2d in the main text).

\section{Error-transparency drive}
In order to achieve error-transparency, we must modify the static interaction Hamiltonian
\begin{equation}
    H_0 = \opad\opa\left(\chi_e\proj{e} + \chi_f\proj{f}\right),
\end{equation}
so that $\chi_e = \chi_f$. We do so by adding a drive at frequency $\omega_\text{ET} = \omega_{eh} - \omega_c + \delta$, where $\omega_{eh}$ is the ancilla transition frequency between $\ket{e}$ and the third excited level $\ket{h}$, $\omega_c$ is the cavity resonance frequency, and $\delta/2\pi \approx 10\unit{MHz}$ is a detuning parameter. This drive addresses the sideband transition $\ket{e,n}\leftrightarrow\ket{h,n-1}$. In the appropriate rotating frame, the resulting Hamiltonian can be written as
\begin{equation}
    H_\text{ET} = g\left(e^{i\delta t}\opa\dyad{h}{e} + e^{-i\delta t}\opad\dyad{e}{h}\right),
\end{equation}
with $g$ the sideband driving rate.
When $\delta \gg g$, we can replace this with an approximate time-independent Hamiltonian
\begin{align}
    H_\text{ET}^\text{eff} &= \frac{g^2}{\delta}\comm{\opa\dyad{h}{e}}{\opad\dyad{e}{h}}\\
    &= \frac{g^2}{\delta}\opad\opa\left(\proj{e} - \proj{h}\right) + \frac{g^2}{\delta}\proj{h}.
\end{align}
Ignoring the terms involving $\ket{h}$, we can write the total effective interaction as 
\begin{align}
    H_0 + H_\text{ET}^\text{eff} = \opad\opa\left(\left(\chi_e + \frac{g^2}{\delta}\right)\proj{e} + \chi_f\proj{f}\right).
\end{align}
For a given value of $g$ (if large enough), one can specify the corresponding value of $\delta$ such that $\frac{g^2}{\delta} = \chi_f - \chi_e$, and therefore the net Hamiltonian becomes
\begin{equation}
H_0 + H_\text{ET}^\text{eff} = \chi_f\opad\opa\left(\proj{e} + \proj{f}\right).
\end{equation}
In practice, we choose $g/2\pi\approx 3 \unit{MHz}$. Increased decoherence induced by the drive and the appearance of unwanted transitions prevent us from using a stronger drive. As a consequence, hybridization of $\ket{e}$ and $\ket{h}$ will ensue, resulting in an approximate population mixing of $g^2/\delta^2\approx 10\%$. This is a second-order error, occurring only after ancilla relaxation takes place, and is included in the error analysis below (section S7).

\section{Error budget}
In order to understand which error mechanisms dominate the residual gate infidelity, we create an analytic model. Several types of errors are accounted for, including first and second-order ancilla transitions during the SNAP operation, transitions during the ancilla measurement, cavity transitions, and readout-induced cavity dephasing. The total predicted 2.1\% error per operation is in good agreement with the measured value of 2.4\%. In the diagram in Fig.~\ref{fig:error_diagram}, we can identify which error channels are dominant, and which are negligible. Second-order transitions, such as double decay from $\ket{f}$ to $\ket{g}$, are of small enough probability to not contribute. Cavity decay, however, is quite significant, both during the SNAP operation (0.4\%) and during the ancilla measurement (0.4\%). In principle, this component can be addressed by introducing the cavity error correction as discussed at the end of section~\ref{sec:ft-snap:error-prop}. Therefore, the dominant and concerning unaddressed error components are: readout-induced dephasing (cross-Kerr, 0.5\%), decoherence-induced back-action (0.3\%), ancilla relaxation during the ancilla measurement (especially from $\ket{e}$, 0.2\%), and hybridization induced by the error-transparency drive, resulting in population of $\ket{h}$ (0.3\%).

\begin{figure*}
    \centering
    \includegraphics[width=\textwidth]{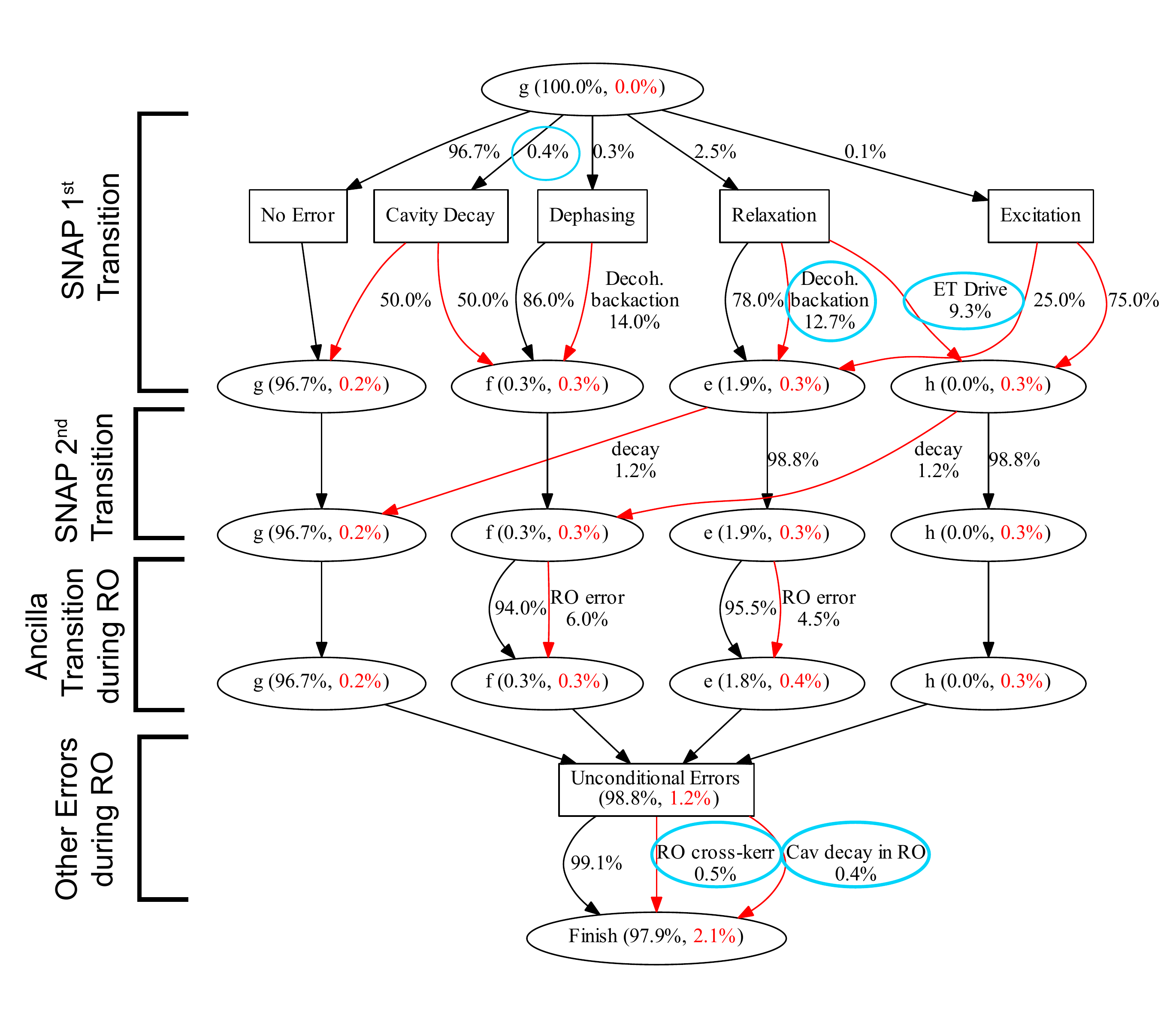}
    \caption{\textbf{Graph of possible error trajectories in the gate protocol.} At each node on the tree, the pair of numbers are probabilities. The first of these is the probability of being in the labelled state and having the logical qubit retain its coherence. The second (in red) is the probability of being in the labelled state and having the logical qubit dephased. The sum of all the numbers in a given row of nodes should be 100\%. Paths between nodes are colored red if the logical qubit is effectively dephased. In the non-error-corrected protocol ($S_\text{NC}$), all paths except the ``No Error'' case would lead to loss of logical qubit coherence, resulting in a predicted fidelity of 96.7\%. The first layer indicates single errors occurring during the SNAP operation. The second layer accounts for second-order double ancilla transition events. The third layer accounts for ancilla relaxation during the readout. The final layer accounts for readout errors that do not depend on the ancilla state. Circled in blue are the primary contributions to the final total, in order, cavity loss ($\sim\!0.8\%$), readout cross-Kerr ($\sim\!0.5\%$), decoherence-induced back-action due to path-independence violation ($\sim\!0.3\%$), relaxation to the third excited state $\ket{h}$ from hybridization induced by the error-transparency drive ($\sim\!0.3\%$) and relaxation from $\ket{e}$ to $\ket{g}$ during readout ($\sim\!0.2\%$).\label{fig:error_diagram}}
\end{figure*}

\begin{figure*}
    \centering
    \includegraphics[height=.45\textheight]{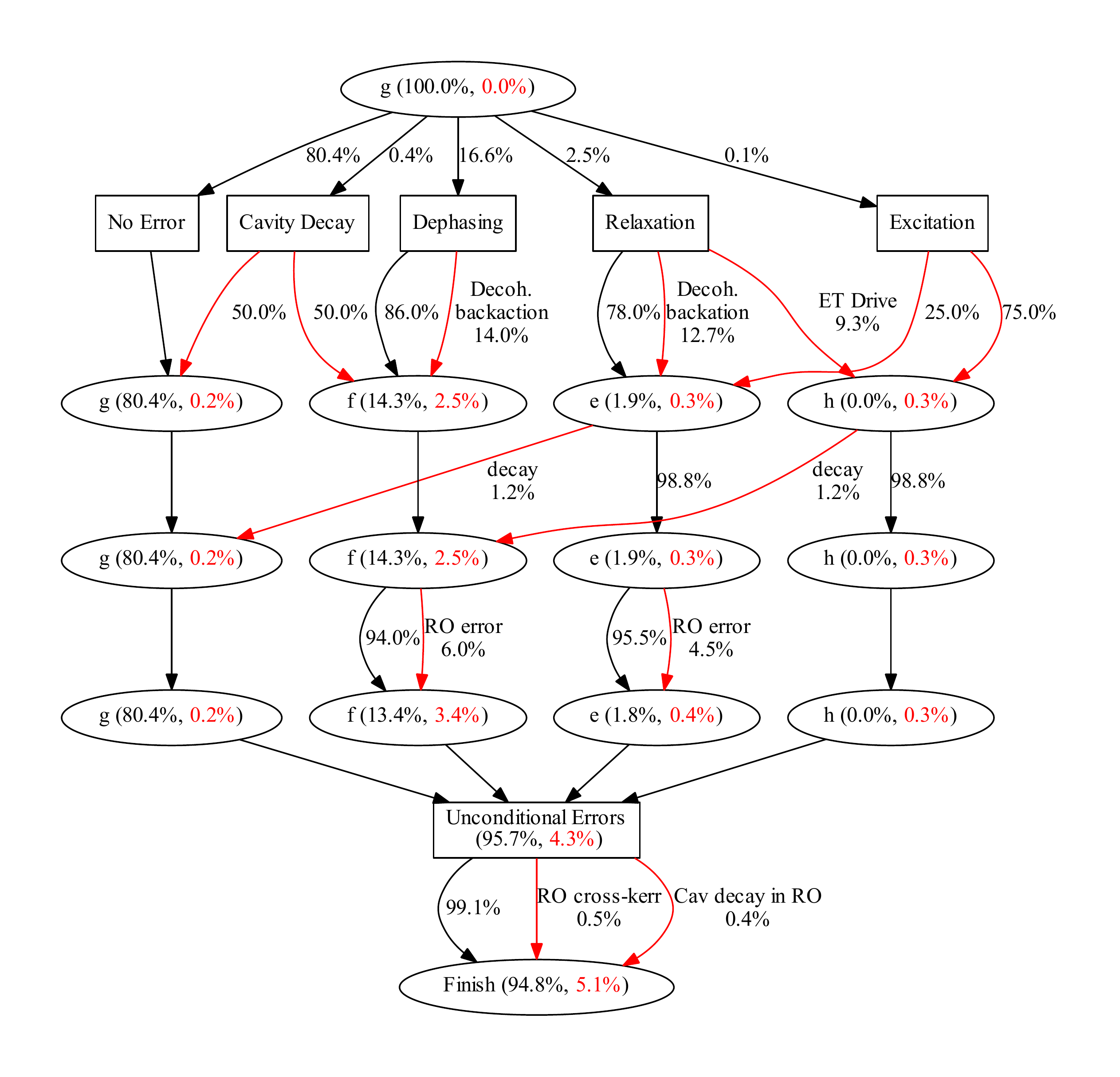}
    \includegraphics[height=.45\textheight]{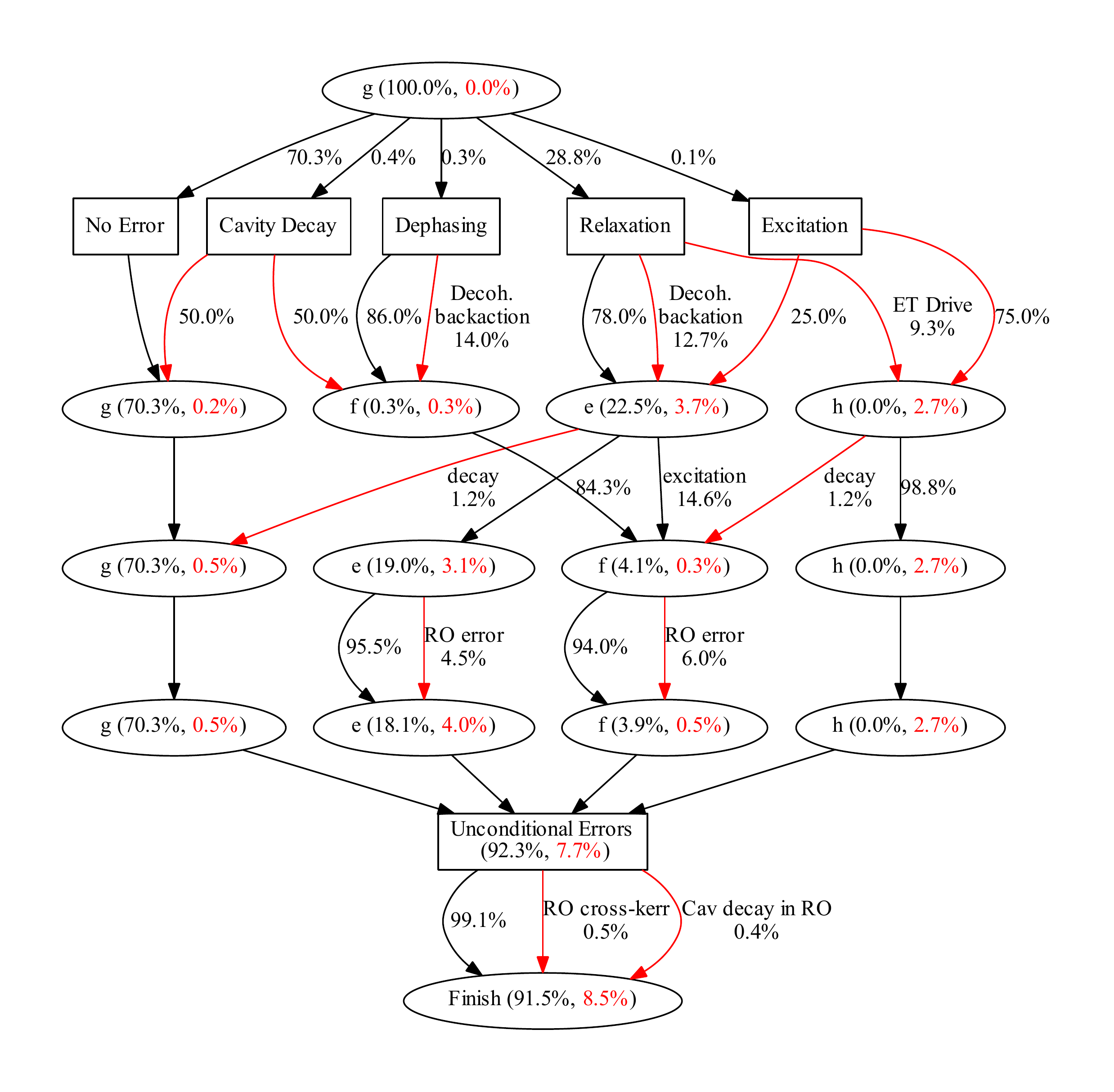}
    \caption{\textbf{Graph of error trajectories in the gate protocol with injected decoherence.} The graphs are shown for the case of maximal induced $g\f$ dephasing (top) and $e\f$ relaxation (bottom) errors.}
\end{figure*}

\begin{table*}[ht]
    \centering
    \ifarxiv\else
    \rowcolors{2}{gray!25}{white}
    \fi
    \begin{tabular}{l l l l l l}
        Parameter Name & Hamiltonian/Liouvillian Term & Quoted quantity & value\\
        \hline
        Transmon frequency & $\omega_{ge}\opbd\opb$ & $\omega_{ge}/2\pi$ &
        $4.2\unit{GHz}$ \\
        Cavity frequency & $\omega_c\opad\opa$ & $\omega_c/2\pi$ &
        $4.5 \unit{GHz}$\\
        Readout frequency & $\omega_\text{RO}r^\dagger r$ & $\omega_\text{RO}/2\pi$ &
        $9.33\unit{GHz}$\\
        Dispersive shift ($\ket{e}$) & $\chi_{e} \opad\opa\proj{e}$ & $\chi_e/2\pi$ &
        $-0.9 \unit{MHz}$\\
        Dispersive shift ($\ket{f}$) & $\chi_{f} \opad\opa\proj{f}$ & $\chi_f/2\pi$ &
        $-1.2 \unit{MHz}$\\
        Transmon anharmonicity & $\frac{\alpha}{2}(\opbd)^2 \opb^2$ & $\alpha/2\pi$ &
        $-137 \unit{MHz}$\\
        Cavity anharmonicity & $\frac{K}{2}(\opad)^2 \opa^2$ & $K/2\pi$ &
        $-2.2\unit{kHz}$ \\
        Transmon $e\rightarrow g$ relaxation& $\frac{1}{T_1^{ge}} D[\dyad{g}{e}]$ & $T_1^{ge}$ &
        $50 \mu s$\\
        Transmon $f\rightarrow e$ relaxation & $\frac{1}{T_1^{e\f}} D[\dyad{e}{f}]$ & $T_1^{e\f}$ &
        $47 \mu s$\\
        Transmon $ge$ dephasing & $\frac{1}{T_\phi^{ge}} D[\opbd\opb]$ & $T_\phi^{ge}$ &
        $>200 \mu s$\\
        Transmon $g\f$ dephasing & $\frac{1}{T_\phi^{g\f}} D[\opbd\opb]$ & $T_\phi^{g\f}$ &
        $>40 \mu s$\\
        Cavity relaxation & $\frac{1}{T_1^c} D[\opa]$ & $T_1^c$ &
        $1.0 \unit{ms}$\\
        Transmon thermal population & $\frac{\bar{n}}{T_1^{ge}}D[\dyad{e}{g}]$ & $\bar{n}$ &
        $0.004$
    \end{tabular}
    \caption{\textbf{System Parameters}}
    \label{tab:parameters}
\end{table*}

\section{Experimental setup}
For a detailed description of the experimental setup, see Ref. \subcite{Rosenblum2018Fault-tolerantError.}. The measured values for the system parameters can be found in table~SI.
\end{linenumbers}
\clearpage

\ifarxiv\else
\printbibliography
\end{refsection}
\fi

\end{document}